\documentclass {elsart}
\usepackage{graphicx,bm,amsmath}
\begin{document}
\begin{frontmatter}
\title{Two-step flux penetration in layered antiferromagnetic superconductor}
\author{T. Krzyszto\'n}
\address {Institute of Low Temperature and Structure Research, Polish Academy of
Sciences, 50-950 Wroc\l{}aw, P.O.Box 1410, Poland}

\begin{abstract}
A layered antiferromagnetic superconductor in the mixed state may
posses magnetic domains created along the Josephson vortices. This
may happen when an external magnetic field is strong enough to
flip over magnetic moments, lying in the phase core of the
Josephson vortex, from their ground state configuration. The
formation of the domain structure of the vortices modifies the
surface energy barrier of the superconductor. During this process
the entrance of the flux is stopped and a newly created state
exhibits perfect shielding. Such behavior should be visible as a
plateau on the dependence of flux density as a function of the
external magnetic field. The end of the plateau determines the
critical field, which we call the second critical field for flux
penetration.
\end{abstract}
\begin{keyword}
High-$T_C$ superconductivity\sep mixed state\sep magnetic
superconductors \PACS 74.60.Ec \sep 74.72.-h
\end{keyword}
\end{frontmatter}
\section*{INTRODUCTION}
Among classical magnetic superconductors there are three groups of
cluster compounds, REMo$_{6}$S$_{8}$, REMo$_{6}$Se$_{8}$, and
RERh$_{4}$B$_{4}$( RE=rare-earth) which have been the primary
systems for study of the interplay between superconductivity and
long-range magnetic order~\cite{Ternary,Maple95}. Although good
quality single crystals of these materials have been available and
measured for a long time a very interesting phenomenon was
recently discovered in DyMo$_{6}$S$_{8}$  due to very carefully
conducted experiment~\cite{Rogacki2001}. This phenomenon predicted
in~\cite{Krzy84} and named two-step flux penetration was
previously observed solely on ($bct$)
ErRh$_{4}$B$_{4}$~\cite{Iwasaki86}. The present work is inspired
by this discovery and the hope that the same behavior could
possibly be observed in some of the layered superconducting
structures. The specific feature caused by the long
antiferromagnetic order in the mixed state of a superconductor is
the possibility of creation of the spin-flop (SF)(or metamagnetic)
domains along the vortices~\cite{Krzy80}. This is easy to
understand by taking two sublattices antiferromagnet as an
example. An infinitesimal magnetic field applied perpendicular to
the easy axis makes the ground antiferromagnetic (AF) state
unstable against the phase transformation to the canted phase
(SF). On the contrary, if the magnetic field is applied parallel
to the easy axis the antiferromagnetic configuration is stable up
to the thermodynamic critical field $H_{T}$. When the field is
further increased a canted phase develops in the system. Let us
assume that in an antiferromagnetic superconductor the lower
critical field fulfils the relation $H_{c1}<\frac{1}{2}H_{T}$ and
that the external field, $H_{c1}<H<\frac{1}{2}H_{T}$ , is applied
parallel to the easy axis. Then the superconducting vortices
appear in the ground antiferromagnetic state. When the field is
increased above $H_{pl}$ (see Fig.\ref{Fig3}) approximately equal
to $\frac {1}{2}H_{T}$ the phase transition to the canted phase
originates in the vortex core. The spatial distribution of the
field around the vortex is a decreasing function of the distance
from its center. Hence the magnetic field intensity in the
neighborhood of the core is less then $H_{T}$. Therefore, the rest
of the vortex remains in the antiferromagnetic configuration. The
radius of the SF domain grows as the field is increased. The above
considerations apply to the classical superconducting Chevrel
phases as well as to the high $T_{c}$ superconductors, where
antiferromagnetic order is produced by the regular lattice of RE
ions occupying isolating layers.
\begin{figure}[!htb]
\begin{center}
\includegraphics*[width=0.6\textwidth]{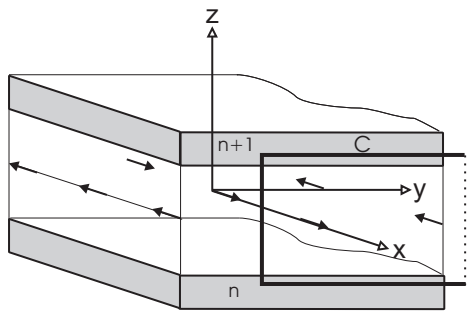}
\end{center}
\caption{Schematic drawing of a piece of the layered
superconductor. The shaded areas (n,n+1) represent superconducting
layers. The bold arrows represent magnetic moments of RE ions
lying in the isolating layers. The axes of the reference frame are
shown.} \label{Fig1}
\end{figure}
In this paper we consider the structure shown on Fig.~\ref{Fig1}
that we believe simulates a real structure of many
antiferromagnetic layered superconductors.  A good candidate to
show the above behavior should possess the isolating layers with
the magnetic moments of RE ions running parallel and antiparallel
to the direction (easy axis) lying in the $ab$ plane. A typical
example of such system is ErBa$_{2}$Cu$_{3}$O$_{7}$. This compound
has tetragonal unit cell with small orthorombic distortion in the
$ab$ plane. The Er ions form two sublattices antiferromagnetic
structure of magnetic moments laying parallel and antiparallel to
the $b$ direction \cite{Zaretsky}. Another example may be Er
nickel boride-carbides ~\cite{Sinha95,Szymczak95,Eskildsen2001}.
The layered  structure of RE nickel boride-carbides is reminiscent
of that of the high-$T_c$ oxide superconductors and  consists of
RE-carbon layers separated by Ni$_{2}$B$_{2}$ sheets.
\section*{BASIC EQUATIONS}
We start description of our problem in terms of the
Lawrence-Doniach energy functional. In this approach a layered
superconductor is described by the superconducting planes with the
interlayer distance d, as shown on Fig.~\ref{Fig1}. The
antiferromagnetic subsystem consisting of RE ions is confined to
the insulating layers. The magnetic moments are running parallel
and antiparallel to the x-axis (easy axis). The Lawrence-Doniach
functional is obtained from the standard Ginzburg-Landau  energy
by discretization of the kinetic energy in the z-direction.
\begin{eqnarray}
F_{S}=\int {d^2}r d\sum_{n}\Bigg[\frac{\hbar ^{2}}{2m} \left|
\left( -i\mathbf{\nabla_{(2)}}+\frac{2ie}{\hbar
}\mathbf{A_{(2)}}\right)\Psi_n \right|^{2}+a \left| \Psi_n
\right|^{2}+\frac{1}{2}b \left| \Psi_n \right|^{4} \nonumber \\
+\frac{\hbar^{2}}{2\mathcal{M}d^{2}}\left|\Psi_{n+1} \exp
\left(\frac{2ei}{\hbar} \int_{nd}^{(n+1)d} A_z dz \right)-\Psi_n
\right|^2 \Bigg]   \label{eq1}
\end{eqnarray}
The quantity $\hbar, e,m,$ denote Planck constant, charge of the
electron and mass of the current carrier in the $ab$ plane,
whereas $\mathcal{M}$ denotes mass of the current carrier in the
z-direction. The antiferromagnetic two sublattices subsystem with
single ion anisotropy is  described with the following energy
density functional
\begin{equation}
f_{M}=\sum_{n}
\Big\{J\mathbf{M}_{1n}\cdot\mathbf{M}_{2n}+K\sum\limits_{i=1}^{2}\left(
M_{in}^{x}\right)^{2}-\left| \gamma
\right|\sum\limits_{i=1}^{2}\sum\limits_{j=x,y,z}( \mathbf{\nabla
}M_{in}^{j}) ^{2}\Big\} \text{.} \label{eq2}
\end{equation}
where $\mathbf{M}_{n}=\mathbf{M}_{1n}+\mathbf{M}_{2n}$ is the sum
of the magnetization vectors of the sublattices in the n-th
insulating layer, $M_{in}^{x}$ is the component of the
magnetization sublattice vector along the anisotropy axis in the
n-th layer, $J$ denotes the exchange constant between two
sublattices, $K$ is the single ion anisotropy constant,
$\sqrt{\left| \gamma \right|}$ is the magnetic stiffness length,
and\ $M_0=\left| \mathbf{M}_{1n}\right| =\left|
\mathbf{M}_{2n}\right| $. Since in the following we analyze the
phenomena with characteristic length-scales much larger then the
interatomic distance it is justified to omit the gradient term in
$f_{M}$. The components of the total magnetization vector
$\mathbf{M}$ have the following form in both sublattices:
$M_{iy}=M_{0}\sin\theta_{i},~M_{ix}=M_{0}\cos\theta_{i}$, where
$\theta _{i}$  (canted spin angle) is the angle between the
magnetization in the sublattice and the external magnetic field
directed along the $x$-axis. The AF $(\theta _{1}=0,\theta
_{2}=\pi )$ and SF phases $(\theta_{1}=-\theta _{2}=\theta )$ are
in thermodynamic equilibrium in an applied field equal to the
thermodynamic critical field~\cite{KKTel}
\begin{equation}
H_{T}=M_{0}[K(J-K)]^{1/2} \text{.} \label{eq3}
\end{equation}
The canted spin angle of the SF phase is then expressed as
\begin{equation}
\cos \theta=\frac {KM_{0}}{H_{T}}  \text{.} \label{eq4}
\end{equation}
Finally we add the magnetic field energy to obtain the free energy
of the entire system
\begin{equation}
F=F_{s}+ \int \Big\{ f_{M}+\frac{\mu _{0}}{2} \mathbf{H}^{2}
\Big\} dV \text{.}\label{eq5}
\end{equation}
According to experiments the antiferromagnetic order is very weak
affected by the presence of superconductivity, then it is
reasonable to neglect the effect of superconductivity on the
exchange interaction in $F$. Instead we introduce electromagnetic
coupling between the magnetic and superconducting subsystem.  This
means that both order parameters $\Psi_n $ and $\mathbf{M}$ are
coupled through the vector potential $\mathbf{A}$
\begin{equation}
\mathbf{B}=\mathrm{rot}\mathbf{A}=\mu
_{0}\mathbf{H}+\mathbf{M}\text{,}\label{eq6}
\end{equation}
\begin{equation}
\bm{j}_s=\mathrm{rot}\mathbf{H} \label{eq7} \text{,}
\end{equation}
where $\mathbf{B}\ $ is the vector of the magnetic flux density
(magnetic induction) and $\mathbf{H}\ $ is the vector of the
thermodynamic magnetic field intensity. The functional (\ref{eq5})
can be treated in the London approximation by assuming a constant
modulus $\Psi_n$ within the planes and allowing only for phase (
$\varphi_{n}$) degree of freedom. The equilibrium conditions of
the whole system can be obtained via minimization the Gibbs free
energy functional
$G=F-\displaystyle\int(\mathbf{B}\mathbf{H}_{0})dV$. Performing
this task with respect to vector potential $\mathbf{A}$ and
$\varphi_{n}$ provides us with the fundamental equations for
currents and phases.
\begin{equation}
\sum_{n}
\frac{d}{\lambda^2}\Big(\frac{\phi_{0}}{2\pi}\mathbf{\nabla_{(2)}}\varphi_{n}
-\mathbf{A_{(2)}}\Big)\delta(z-nd)=\mu_0
\mathbf{j_{(2)}}=\mathrm{rot}_{(2)}\mathbf{(B-M)} \label{eq8}
\end{equation}
\begin{equation}
\sum_{n}\Big(\frac{\phi_{0}}{2\pi}\frac{1}{\lambda^2_{j}d}\sin\chi_{n+1,n}\Big)
\Theta(z-dn)\Theta[d(n+1)-z]=\mu_0
j_{z}=\mathrm{rot}_{z}\mathbf{(B-M)} \label{eq9}
\end{equation}
\begin{equation}
\mathbf{\nabla_{(2)}}\Big(\mathbf{\nabla_{(2)}}\varphi_{n}-\frac{2\pi}{\phi_{0}}\mathbf{A_{(2)}}
\Big)=\frac{1}{{r_{j}}^2}\Big(
\sin\chi_{n+1,n}-\sin\chi_{n,n-1}\Big) \label{eq10}
\end{equation}
where $\delta(z-nd)$ is the Dirac delta function, $\Theta(z-dn)$
Heaviside step function, $\lambda$ denotes London penetration
depth in the superconducting plane,
$\lambda_{j}=\lambda\sqrt{\mathcal{M}/m}$,
$r_j=d\sqrt{\mathcal{M}/m}$ and $\chi_{n+1,n}=
\varphi_{n+1}-\varphi_{n}+\displaystyle\frac{2ei}{\hbar}
\int_{nd}^{(n+1)d} A_z dz$ is the gauge invariant phase
difference. In the following we shall investigate the problem of a
single vortex line lying parallel to the Josephson coupled
superconducting layers, separated by the insulating
antiferromagnetic layers.
\begin{figure}[!htb]
\begin{center}
\includegraphics*[width=0.6\textwidth]{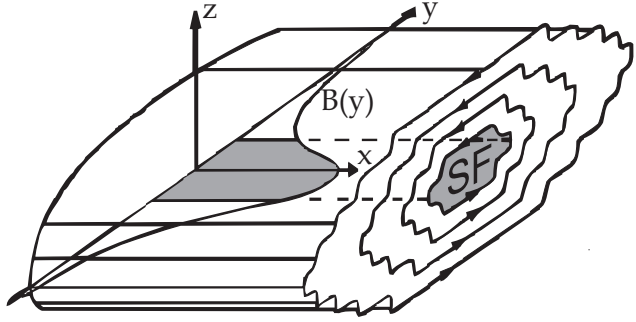}
\end{center}
\caption{Single Josephson vortex lying in the $ab$ plane along the
$\hat{x}$-axis. The shaded area shows induced SF domain along the
phase core.} \label{Fig2}
\end{figure}
\section*{SINGLE JOSEPHSON VORTEX}
We assume that the vortex center, located in the central $n=0$
layer, is parallel to the $x$-axis. The relation between the
magnetic field of the vortex and the gauge-invariant phase
difference $\chi_{n+1,n}(y)$ of the superconducting wave function
across layers $n$ and $n+1$ can be found by integrating the vector
potential given by equations (\ref{eq8}) and (\ref{eq9}) around a
rectangular, semi-infinite contour C, as shown in Fig.~\ref{Fig1}.
This contour, parallel to the $yz$ plane, is located apart from
the central junction $n=0$, where nonlinearities of the phase
difference must be taken into account. The magnetic flux within
this contour in given as. \[ \Phi(y)=d\int_{y}^{\infty}d y'
B(y',z)=\oint_{C}\mathbf{A}d\bm{l}\] Differentiating the result
with respect to $y$ one obtains
\begin{equation}
\mathbf{B}=
\lambda^2\frac{\partial^{2}\big(\mathbf{B-M}\big)}{\partial z^2}
+\lambda_{j}^2\frac{\partial^{2}\big(\mathbf{B-M}\big)}{\partial
y^2} \label{eq11}
\end{equation}
The above equation shows that in the Josephson vortex the
screening currents vanish on a length scale $\lambda_{j}$ along
$y$ axis, and a scale $\lambda$ along $z$ direction. On these
scales the Josephson and Abrikosov vortices in an anisotropic
superconductor are roughly equivalent apart from small corrections
in the current flow, Fig.~\ref{Fig2}, and the magnetic field
pattern due to the layered structure. But in contrast to the
Abrikosov vortex, where the large current flow near the core leads
to complete supression of the order parameter, the supression of
the order parameter in the superconducting layers is only weak in
the so called phase core of the Josephson vortex. Within the
distances $r_j$ along $y$,   and $d$ along $z$ we have to take
into account the nonlinearity and discretness of Eq.~(\ref{eq10}).
On these length scales the phase is changing rapidly and the
current density $j_z$ reaches its maximum value
$j_c=\displaystyle\frac{\phi_0}{2\pi\mu_0\lambda_{j}^2d}$. In the
region of the phase core London model fails.

To make the problem simpler we assume that the magnetization in
the isolated vortex is constant across the SF phase domain. Thus,
we can write
\begin{equation}
|\mathbf{M}| =\left\{
\begin{array}{ccc}
M & ~~\text{if} & \rho\leq \rho_{m} \\
0 & ~~\text{if} & \rho>\rho_{m}
\end{array}
\right. \text{,} \label{eq12}
\end{equation}
where $\rho_{m}$ is dimensionless radius of the $SF$ domain in the
cylindrical reference frame
~$x=x~;~y=\lambda_j\rho\sin\theta~;~z=\lambda\rho\sin\theta$. Then
the solution of Eq.~(\ref{eq11}) for a single  Josephson vortex is
given by the modified Bessel functions $K_{0}$ and $I_0$
\begin{eqnarray}
b_\mathrm{ SF} &=&  C_{1}K_{0}\left(\rho\right) +C_{2}I_{0}\left(\rho\right)~\text{,~for}~\rho_j
<\rho\leq \rho_{m}  \nonumber \\
b_\mathrm{ AF} &=&
C_{3}K_{0}\left(\rho\right)~\text{,~for}~\rho>\rho_{m} \text{,}
\label{eq13}
\end{eqnarray}
($\rho_j $ denotes the dimensionless phase coherence length ) with
the following boundary conditions:
\begin{eqnarray}
b_{\mathrm{SF}}\left(\rho_m\right) &=& \mu_{0}H_{T}+M = B_{T}\nonumber\\
b_{\mathrm{AF}}\left(\rho_{m}\right) &=&\mu_{0} H_{T}\text{.}
\label{eq14}
\end{eqnarray}
These conditions, together with the flux quantization condition,
are used to calculate the arbitrary constants in Eq.~(\ref{eq13}).
\begin{eqnarray}
&&C_{1}=\frac{\displaystyle B_{T} \rho_m I_{1}\left(
\rho_m\right)-\left[\mu_0 H_{T}\rho_m\frac{K_{1}\left( \rho_m
\right)} {K_{0}\left( \rho_m\right)}-\frac{\varphi _{0}}{2\pi
\lambda\lambda_j}\right] I_{0}\left(
\rho_m\right)}{\displaystyle\rho_{m} K_{1} \left( \rho_m\right)
I_{0}\left( \rho_m\right) -I_{0}
\left( \rho_m\right) +\rho_m K_{0}\left( \rho_m\right)I_{1}\left( \rho_m\right)}\nonumber \\
&&C_{2}=\frac{\displaystyle{B_{T}\left[ \rho_m K_{1}\left( \rho_m
\right) -1\right] +\left[\mu_0 H_{T}\rho_m\frac{K_{1}\left( \rho_m
\right)}{K_{0}\left( \rho_m\right)}-\frac{\varphi
_{0}}{2\pi\lambda\lambda_j}\right] K_{0}\left(
\rho_m\right)}}{\displaystyle\rho_m K_{1}\left( \rho_m\right)
I_{0}\left( \rho_m\right) -
I_{0}\left(\rho_m\right) +\rho_m K_{0}\left(\rho_m\right)I_{1}\left(\rho_m\right)}\nonumber \\
&&C_{3}=\frac{\mu_0 H_{T}}{K_{0}\left(\displaystyle
\rho_m\right)}\text{.} \label{eq15}
\end{eqnarray}
Finally we write free energy of the isolated vortex
\begin{eqnarray}
\varepsilon &=&\frac{\lambda_{j}\lambda}{2\mu_0}\oint_{\sigma _{1}}d \bm{\sigma}
\left\{\left[\textbf{b}_\mathrm{SF}\left(\bm{r}\right)-\textbf{M}\right]\times \mathrm{rot}
\textbf{b}_\mathrm{ SF}\left(\bm{r}\right)\right\}\nonumber\\
&+&\frac{\lambda_{j}\lambda}{2\mu_0}\oint_{\sigma_{2}}d
\bm{\sigma}\left\{ \textbf{b}_\mathrm{AF}\left(\bm{r}\right)\times
\mathrm{rot}\textbf{b}_\mathrm{AF}\left(\bm{r}\right)\right\}\text{,}
\label{eq16}
\end{eqnarray}
where
$\bm{r}=(\displaystyle{\frac{y}{\lambda_j},\frac{z}{\lambda}})$ is
the position of the vortex line, $\sigma_1$ denotes the surface of
the phase core, and $\sigma_2$ the surface of the SF domain
respectively. The integrals in Eq.~(\ref{eq16}) performed as line
integrals along the contours of the cross sections of the
appropriate  surfaces  give $\varepsilon _{1}$ - the line tension
of the vortex. The minimum of $\varepsilon _{1}$ with respect to
$\rho_m$ determines
\begin{equation}
\rho_{m }^{2}=\frac{5\phi _{0}}{8\pi\lambda\lambda
_{j}B_{T}}\text{.} \label{eq17}
\end{equation}
\section*{FREE ENERGY OF THE LATTICE}
Equation (\ref{eq11}), in the new coordinates, can be rewritten
for the lattice of vortices in the following way:
\begin{equation}
\mathbf{B}+\mathrm{rot}\mathrm{rot}\mathbf{B}=\frac{\phi
_{0}}{\lambda\lambda _{j}}\sum_m\delta(\bm{r}-\bm{r}_m)\text{,}
\label{eq18}
\end{equation}
where $r_m$ specify the positions of the phase cores of the
vortices. The solution of Eq.~(\ref{eq18}) is then a superposition
\[\mathbf{B}(\bm{r})=\sum_m \mathbf{B}_m(\bm{r}-\bm{r}_m)\]
of the solutions $\mathbf{B}_m(\bm{r}-\bm{r}_m)$ of isolated
vortices at points $\bm{r}_m$. The free energy of the system can
thus be  written as
\begin{equation}
F=\frac{\lambda\lambda_j}{2\mu_0}\oint_{\sigma }d
\bm{\sigma}(\mathbf{B}\times\mathrm{rot}\mathbf{B}) \label{eq19}
\end{equation}
The above symbolic surface integral is taken over the surfaces of
the phase cores and surfaces of the SF domains. The energy of the
Meissner state is chosen as zero of the energy scale. Again, when
the surface integrals are replaced by contour ones over
appropriate cross sections we get line energy of the system. This,
in turn, multiplied by vortex density $n$ gives $f$-free energy
density of the system. After some transformations one can derive
the following formula
\begin{equation}
f=n\varepsilon _{1} + n\phi_0 H_T (\ln \beta)^{-1} \sum_m
K_0(r_m)\text{,} \label{eq20}
\end{equation}
\[ \beta=\sqrt{\frac{\displaystyle\pi\lambda\lambda_j B_T}{\displaystyle\phi _{0}}},\]
here the sum is over all vortices excluding the one in the origin,
and $r_m$ denotes the distance of a vortex from the origin. The
lattice sum may now be replaced by integral in the $yz$-plane over
a smoothed vortex density, excluding the area $n^{-1}$ associated
with the single flux line in the origin. The free energy density
then reduces to
\begin{equation}
f=n\varepsilon _{1}+B^2\Big(\frac{\displaystyle H_T}{\displaystyle
B_T}\Big)\Big(\frac{\displaystyle\beta}{\displaystyle\ln\beta}\Big)+B
\frac{\displaystyle{H_T}}{\displaystyle
4\ln\beta}\sqrt{\frac{\displaystyle 4\lambda_j}
{\displaystyle27\lambda}}\ln\Big(\frac{\displaystyle{a}}{\displaystyle\sqrt{\lambda\lambda_j}}
\Big) \label{eq21}
\end{equation}
\[ \Big(\frac{\displaystyle a}{\displaystyle\sqrt{\lambda\lambda_j}}\Big)^2=
\frac{\displaystyle 1}{\displaystyle\beta^2}\Big(\frac{
\displaystyle B_T} {\displaystyle B}\Big)\sqrt{\frac{\displaystyle
4\lambda_j}{\displaystyle 27\lambda}},\] here $a=|\bm{a}_1|$
denotes the length of the basal vector of the nonequilateral
triangular unit cell, and $2|\bm{a}_2|=a\sqrt{1+\tan^2\alpha}$ (
$\alpha $ is the angle between both vectors),~
$\tan\alpha=\sqrt{\frac{3\lambda}{\lambda_j}}$~\cite{Kogan81}. To
determine the equilibrium state it is necessary to minimize the
Gibbs free energy density with respect to magnetic induction. The
result yields an implicit equation for the constitutive relation
$B=B(H)$
\begin{equation}
H-\frac{\varepsilon _{1}}{\phi_0}=B\Big(\frac{\displaystyle
H_T}{\displaystyle B_T}\Big)\Big(\frac{\displaystyle 2
\beta}{\displaystyle\ln\beta}\Big)+\frac{\displaystyle{H_T}}{\displaystyle
4\ln\beta}\sqrt{\frac{\displaystyle 4\lambda_j}
{\displaystyle27\lambda}}\ln\Big(\frac{\displaystyle{a}}{\displaystyle\sqrt{\lambda\lambda_j}}
\Big) \label{eq22}
\end{equation}
\section*{FLUX PENETRATION}
Consider semi-infinite specimen in the half space $y\geq 0$, the
vortex and the external magnetic field running parallel to the
surface in the $x$ direction. The presence of a surface of the
superconductor leads to a distortion of the field and current of
any vortex located within a distance of the order of penetration
depth from the surface. To fulfill the requirement that the
currents cannot flow across the surface of the superconductor we
need to introduce an image vortex, with  vorticity opposite to the
real one. Both vortices, direct and image, interact as real ones
except that the interaction is attractive. In the low flux density
regime, Clem \cite{ClemLT} has shown that there exist two regions:
a vortex-free region of the width $y_{ff}$ near the surface of the
sample, and a constant flux density region for $y>y_{ff}$. Within
the vortex-free area one can introduce the locally averaged
magnetic field $B_{M}$ which is  a linear superposition of the
Meissner screening field, the averaged direct vortices flux
density exponentially decreasing towards the surface from its
interior value $B$ at $y=y_{ff}$, and averaged image vortices flux
density. In our problem the $x$ component of this superposition
can be approximated by
\begin{equation}
B_{M}=B\cosh \left(\frac{y_{ff}-y}{\lambda_j }\right)\text{.}
\label{eq23}
\end{equation}
The boundary condition $B_M(0)=\mu_0 H_0$ determines the thickness
of the vortex-free region
\begin{equation}
y_{ff}=\lambda_j \cosh^{-1}
\left(\frac{\mu_0H_{0}}{B}\right)\text{.} \label{eq24}
\end{equation}
We assume that the test vortex line is lying within vortex free
region at a point $\bm{r}=(\displaystyle{\frac{y}{\lambda_j},0})$,
and its image at $\bm{r}=(\displaystyle{-\frac{y}{\lambda_j},0})$
outside the superconductor. Now the local field of the test vortex
can be understood as a superposition of the following fields
\begin{eqnarray}
\mathbf{B}_\mathrm{ SF}&=&\mathbf{b}_\mathrm{ SF}\left(\bm{r}\right)-\mathbf{b}_\mathrm{ AF}
\left(2\bm{r}\right) +\hat{x}B_{M}\left(\bm{r}_{ff}-\bm{r}\right) \nonumber \\
\mathbf{B}_\mathrm{ AF}&=&\mathbf{b}_\mathrm{
AF}\left(\bm{r}\right)-\mathbf{b}_\mathrm{ AF}\left(2\bm{r}\right)
+\hat{x}B_{M}\left(\bm{r}_{ff}-\bm{r}\right)\text{,} \label{eq25}
\end{eqnarray}
where $\bm{r}_{ff}=(\displaystyle{\frac{y_{ff}}{\lambda_j},0})$,
and $\hat{x}$ denotes the unit vector in the $x$ direction. Having
determined the local magnetic field we can write the Gibbs free
energy of the test vortex line as
\begin{eqnarray}
G&=&\frac{\lambda\lambda_j }{2\mu_0}\oint_{\sigma _{1}}d
\bm{\sigma}\left\{\left[ \mathbf{B}_\mathrm{
SF}\left(\bm{r}\right)-2\mu_0\mathbf{H}_{0}-\mathbf{M}\right]
\times\mathrm{rot} \mathbf{B}_\mathrm{SF}\left(\bm{r}\right)\right\}\nonumber\\
&+&\frac{\lambda\lambda_j }{2\mu_0}\oint_{\sigma _{2}}d
\bm{\sigma} \left\{\left[\mathbf{B}_\mathrm{
AF}\left(\bm{r}\right)-2\mu_0\mathbf{H}_{0}\right]
\times\mathrm{rot} \mathbf{B}_\mathrm{AF}\left(\bm{r}\right)\right\}\nonumber \\
&+&\frac{\lambda\lambda_j }{2\mu_0}\oint_{\sigma _{2}}d
\bm{\sigma}
\left\{\hat{x}B_{M}\left(\bm{r}_{ff}-\bm{r}\right)\times\mathrm{rot}
\mathbf{B}_\mathrm{AF}\left(\bm{r}\right)\right\}\text{.}
\label{eq26}
\end{eqnarray}
After some transformations ~\cite{ClemLT,Krzy94} one can obtain
the Gibbs free energy per unit length $\mathcal{G}$
\begin{equation}
\mathcal{G}=\mathcal{G}_{1}+\mathcal{G}'_{1}+\mathcal{G_M}\text{,}
\label{eq27}
\end{equation}
where
\begin{eqnarray}
\mathcal{G}_{1} &=& \varepsilon _{1}-\frac{\lambda\lambda_j\pi}{4\mu_0}D_1b_\mathrm{ AF}
\left(2r\right)\nonumber\\
\mathcal{G}'_{1} &=& -\frac{\lambda\lambda_j\pi}{2\mu_0}D_1\left[b_\mathrm{ AF}
\left(r_{ff}\right)-b_\mathrm{ AF}\left(r_{ff}+r\right)\right]\nonumber\\
\mathcal{G_M} &=& -\frac{\lambda\lambda_j\pi}{2\mu_0}\left[
D_{1}\mu_{0} H_{0} - D_{2}
B_{M}\left(r_{ff}-r\right)\right]\text{,} \label{eq28}
\end{eqnarray}
and
\begin{eqnarray}
D_{1}&=& \left. -\rho_j\frac{d b_\mathrm{ SF}(\rho)}{d\rho}\right|_{\rho=\rho_j}
\left. -\rho_m\frac{d b_\mathrm{ SF}(\rho)}{d\rho}\right|_{\rho=\rho_m}\left.
-\rho_m\frac{d b_\mathrm{ AF}(\rho)}{d\rho}\right|_{\rho=\rho_m}\nonumber\\
D_{2}&=&\left. -\rho_j\frac{d b_\mathrm{
SF}(\rho)}{d\rho}\right|_{\rho=\rho_j}\left. -\rho_m\frac{d
b_\mathrm{ SF}(\rho)}{d\rho}\right|_{\rho=\rho_m}\left.
-2\rho_m\frac{d b_\mathrm{ AF}(\rho)}{d\rho}\right|_{\rho=\rho_m}
\label{eq29}
\end{eqnarray}
$\mathcal{G}_{1}$ describes the interaction of the test vortex
with its image, $\mathcal{G}'_{1}$ is a correction term introduced
by Clem \cite{ClemLT}, and $\mathcal{G_M}$ describes the
interaction energy of the test vortex with the modified Meissner
field. To find the conditions of the vortex entrance and exit, one
has to solve a force balance equation for the test vortex, at the
surface of the sample, and at the edge of the flux-filled area,
respectively. A calculation using $\mathcal{G}_{1}$ and
$\mathcal{G_M}$ alone gives non vanishing force on the test vortex
at $\bm{r}=\bm{r}_{ff}$. However, the force should be zero there,
because $\mathcal{G_M}$ is supposed to account for all the image
vortices. To avoid double counting the image vortex one can
subtract from the self-energy  a contribution of the excess image
fixed at $\bm{r}=-\bm{r}_{ff}$. One can easily check that
$\mathcal{G}'_{1}$ is negligible at the surface of the sample and
has no influence on the conditions of the flux entrance. When the
flux starts to enter the sample, $H_0=H_{en2}(B)$,
\begin{equation}
y_{ff}=y_{en}=\lambda_j \cosh^{-1}\Big(\frac{\mu_0
H_{en2}(B)}{B}\Big)\text{,} \label{eq30}
\end{equation}
and the energy barrier is moved toward the surface within
$\rho_m$. Thus, one can derive from the force balance equation
\begin{equation}
-\frac{D_1}{2D_2}\left.\frac{d b_\mathrm{
AF}(\rho)}{d\rho}\right|_{\rho=\rho_m}=B\sinh\Big(\frac{y_{en}}{\lambda_j}\Big)\text{.}
\label{eq31}
\end{equation}
The left hand side of the above equation gives $H_{en2}(0)=
H_T\beta(2\ln\beta)^{-1}$. This field may be thought as the second
critical field for flux penetration calculated in the single
vortex approximation \cite{Krzy94}. Combining Eqs.~(\ref{eq30})
and (\ref{eq31}) we finally obtain
\begin{equation}
H_{en2}(B)=\sqrt{B^2+\Big(\frac{\mu_0
H_{T}\beta}{2\ln\beta}\Big)^2} \text{.} \label{eq32}
\end{equation}
In the opposite case, when the flux exits the sample, the surface
energy barrier tends to the edge of the flux-filled zone. Similar
considerations as the above show that
\begin{equation}
\mu_0 H_{ex2}(B)\simeq B \text{.} \label{eq33}
\end{equation}
The measure of the height of the energy barrier against flux
entrance is
\[\Delta H_{en}(B)=\left|H_{en2}(B)-H_{eq}(B)\right|,\]
and against flux exit
\[\Delta H_{ex}(B)=\left|H_{eq}(B)-H_{ex2}(B)\right|,\]
where $H_{eq}$ is given by Eq.~(\ref{eq22}).
\section*{DISCUSSION OF THE RESULTS}
\begin{figure}[!htb]
\begin{center}
\includegraphics*[width=0.6\textwidth]{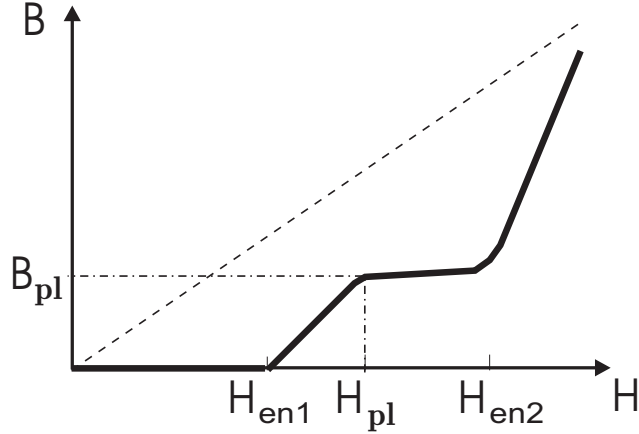}
\end{center}
\caption{Schematic drawing of the magnetization process. $H_{en1}$
denotes the first penetration field for vortices without magnetic
structure, $H_{pl}$ is the applied field which originates SF
transitions inside vortices, and $B_{pl}$ is the corresponding
flux density.  $H_{en2}$ is the entrance field for the vortices
possessing magnetic structure.} \label{Fig3}
\end{figure}

Let us make a short summary of the calculations and visualize the
results on schematic magnetization curve shown in the
Fig.~\ref{Fig3}.
When the external field is not strong enough to create the SF
domains inside vortices, than the magnetization process of the
sample being entirely in the AF phase is as follows. The vortices
without magnetic structure start to enter the specimen at
$H_{en1}$. When the field is increased up to the value $H_{pl}$,
which is of the order of $H_T$, the SF domains are created. Now,
the screening current must redistribute its flow in order to keep
constant the flux carried by the vortex. This feature is easily
seen from Eqs.(\ref{eq13}~-~\ref{eq15}). The redistribution of the
screening current changes the surface energy barrier  preventing
vortices from entering the sample as expressed in
Eq.~(\ref{eq28}). It means that the density of vortices $n$ is
kept constant. Consequently the averaged flux density in the
sample $B=n\varphi_0$ remains constant when the external field is
increased. In Fig.~\ref{Fig3} this feature is visible as a plateau
on the $B(H)$ curve, or alternatively as a second negative slope
on the $M(H)$ curve. The vortices start to penetrate the sample
when the external field reaches the right edge of the plateau. We
call this value, given by Eq.~(\ref{eq32}), second critical field
for flux penetration $H_{en2}$.

To find the thermodynamic critical field $H_{T}$, and then to
calculate $H_{en2}(B)$ the following argumentation is proposed. At
low fields, in the vicinity of the lower critical field $H_{c1}$,
the intensity of the field in the vortex core is
$2H_{c1}$~\cite{ClemCoffey90}. When the external field is
increased the field intensity in the vortex core increases because
of the superposition of the fields of the surrounding vortices.
The field intensity in the core must reach $H_T$ in order to
originate a transition to the SF phase. Thus, taking into account
only the nearest neighbors we can write for the nonunilateral
triangular lattice
\begin{equation}
H_{T}=2H_{c1}+z\frac{\varphi _{0}}{\pi
\lambda\lambda_j\mu_0}\left[K_{0}\left(\frac{a}{\lambda_j}\right)+
2K_{0}\left(\frac{a}{2\lambda_j}\sqrt{\frac{3\lambda}{\lambda_j}}\right)\right]
\text{,} \label{eq34}
\end{equation}
here $a$ corresponds to the value $B_{pl}$ of the flux density for
which the penetration process stops, see Fig.(~\ref{Fig3}). From
the relation $B_{pl} = 2\varphi
_{o}\sqrt{\lambda_j}/(a^{2}\sqrt{3\lambda})$ one can compute $a$,
which in turn may be inserted back into Eq.~(\ref{eq34}). It is
easy to estimate the saturation magnetization $M_0$ taking into
account the volume of the elementary cell. Then, Eqs.~(\ref{eq3})
and (\ref{eq4}) can be used to calculate $M$ in the SF-phase
domain
\begin{equation}
M = 2M_{0}\cos\theta = \frac{2KM_{0}^{2}}{H_{T}}\text{.}
\label{eq35}
\end{equation}
\section*{CONCLUSION}
The layered antiferromagnetic superconductor may reveal below
$T_N$ a very interesting behavior in the magnetic field applied
parallel to the superconducting planes. When the sample is in the
virgin state, initially it magnetizes like ordinary type II
superconductor. Upon the applied magnetic field of intensity equal
to the critical field for flux penetration the sample undergoes a
transformation from the Meissner to the mixed state. Then, the
magnetization may proceed in an unusual way. When the field is
further increased a new state may appear in which vortices
possesses the spin-flop phases created around the cores. We have
assumed that in this new state vortices undergo metamorphosis to
the shape shown in Fig.~\ref{Fig2}. This state is characterized by
the plateau on the magnetization curve, shown in the
Fig.~\ref{Fig3}. It means that the magnetic flux density inside
the sample is unaffected by an increased external field. This
perfect shielding should occur until the applied field reaches
certain value of intensity, we call it second critical field for
flux penetration. Then the vortices possessing magnetic structure
enter into the sample. This phenomenon we named two-step flux
penetration.
\section*{ACKNOWLEDGEMENTS}
The author would like to thank P. Tekiel and K. Rogacki for
helpful discussions. This work was supported by the State
Committee for Scientific Research (KBN) within the Project
\mbox{No. 2 P03B 125 19}.

\end{document}